\documentclass[aip,jcp,amsmath,amssymb,preprint]{revtex4-1}

\usepackage{color}
\usepackage{graphicx}
\usepackage{dcolumn}
\usepackage{bm}
\usepackage[utf8x]{inputenc}
\usepackage[T1]{fontenc}

\newcommand {\Ca} {C$^\alpha$}
\newcommand {\Cb} {C$^\beta$}
\newcommand  {\Eev} {E_\text{ev}}
\newcommand  {\Ehb} {E_\text{hb}}
\newcommand  {\Eloc} {E_\text{loc}}
\newcommand  {\Esc} {E_\text{sc}}
\newcommand  {\Pn}	   {P_\text{n}}
\newcommand  {\kf}      {k_{\text{f}}}
\newcommand  {\ku}      {k_{\text{u}}}
\newcommand  {\ltm}      {\tilde \lambda}
\newcommand  {\lcm}      {\hat \lambda}
\newcommand  {\nhb}      {n_\text{hb}}
\newcommand  {\tcm}      {\tau_\text{cm}}
\newcommand  {\ttm}      {\tau_\text{tm}}

\newcommand   {\ev}[1]          {\langle#1\rangle}

\newcommand*{\citen}[1]{%
  \begingroup
    \romannumeral-`\x 
    \setcitestyle{numbers}%
    \cite{#1}%
  \endgroup   
}

\usepackage{color}

\begin{document}

\title[]{Markov modeling of peptide folding in the presence of protein crowders}

\author{Daniel Nilsson}
\email{daniel.nilsson@thep.lu.se} 
\affiliation{Computational Biology and Biological Physics,
Department of Astronomy and Theoretical Physics, Lund University, S\"olvegatan 14A, SE-223 62 Lund, Sweden}

\author{Sandipan Mohanty}
\email{s.mohanty@fz-juelich.de}
\affiliation{Institute for Advanced Simulation, J\"ulich
Supercomputing Centre, Forschungszentrum J\"ulich, D-52425 J\"ulich, Germany}

\author{Anders Irb\"ack}
\email{anders@thep.lu.se}
\affiliation{Computational Biology and Biological Physics,
Department of Astronomy and Theoretical Physics, Lund University, S\"olvegatan 14A, SE-223 62 Lund, Sweden}

\date{\today}

\begin{abstract}
We use Markov state models (MSMs) to analyze the dynamics 
of a $\beta$-hairpin-forming peptide in Monte Carlo (MC) simulations 
with interacting protein crowders, for two different types of 
crowder proteins [bovine pancreatic trypsin inhibitor (BPTI) and GB1].  
In these systems, at the temperature used, the 
peptide can be folded or unfolded and bound or unbound to 
crowder molecules. Four or five major free-energy minima
can be identified. To estimate the dominant MC relaxation
times of the peptide, we build MSMs using a range of different 
time resolutions or lag times.  
We show that stable relaxation-time estimates can be obtained from 
the MSM eigenfunctions through fits to autocorrelation data. The 
eigenfunctions remain sufficiently accurate to permit stable 
relaxation-time estimation down to small lag times, at which point
simple estimates based on the corresponding eigenvalues have
large systematic uncertainties.  The presence of the crowders 
have a stabilizing effect on the peptide, 
especially with BPTI crowders, which can be attributed 
to a reduced unfolding rate $\ku$, while the folding rate $\kf$ is 
left largely unchanged.  
\end{abstract}


\maketitle

\section{Introduction}

In the crowded interior of living cells, proteins are surrounded by high
concentrations of macromolecules, which leads to a reduction of 
the volume available to a given protein.  
Under such conditions, steric interactions would universally 
favor more compact structures.
A growing body of evidence indicates, 
however, that the effects of macromolecular crowding on properties 
such as protein stability cannot be explained in terms of steric repulsion 
alone.\cite{Guzman:14,Monteith:15,Danielsson:15} 
To understand the role of other interactions, 
in recent years, there have been increasing efforts 
to perform computer simulations with realistic crowder 
molecules,\cite{McGuffee:10,Feig:12,Predeus:12,Macdonald:15,Bille:15,Yu:16,Feig:17,Qin:17} 
rather than hard-sphere crowders. When analyzing 
these large systems, a major challenge lies in 
identifying the main states and dynamical modes, which 
may not be easily anticipated. One possible approach to this problem 
is provided by Markov modeling 
techniques,\cite{Schutte:99,Chodera:07,Buchete:08,Bowman:09,Prinz:11} which 
in recent years have found widespread use in studies of biomolecular 
processes such as folding and binding.\cite{Chodera:14,Noe:17}  
Most of these studies dealt with data from molecular dynamics 
simulations, but the methods are general and can be used on Monte 
Carlo (MC) data as well.   

In this article, we use Markov modeling, along with time-lagged independent component analysis 
(TICA),\cite{Molgedey:94,Naritomi:13,Schwantes:13,PerezHernandez:13}
to analyze data from MC simulations of a test peptide in the presence 
of interacting protein crowders, for two different types of crowder proteins. 
We show that the major free-energy minima 
and slow dynamical modes of these high-dimensional systems 
can be identified in a systematic manner using TICA and Markov state models (MSMs). 
We further show that the dominant MC relaxation times of the peptide can be  
robustly estimated from the constructed MSMs, although simple estimates 
based on the MSM eigenvalues are subject to well-known systematic uncertainties. Our
procedure for relaxation-time estimation uses the MSM eigenfunctions and 
autocorrelation fits, rather than the eigenvalues.
  
As test molecule, we use the $\beta$-hairpin-forming 
GB1m3 peptide.\cite{Fesinmeyer:04} The peptide is simulated in homogeneous 
crowding environments, where either bovine pancreatic trypsin 
inhibitor (BPTI) or the B1 domain of streptococcal protein G (GB1) 
serves as crowding agent. Both these  
proteins are thermally highly stable\cite{Moses:83,Gronenborn:91}
and therefore modeled using a fixed-backbone approximation, whereas
the GB1m3 peptide is free to fold and unfold in the simulations. 
The simulations are conducted using MC dynamics at 
constant temperature. Recently, we studied the same systems 
using MC replica-exchange methods, and found that both BPT1 
and GB1 have a stabilizing effect on GB1m3.\cite{Bille:16} 

\section{Methods}

\subsection{Simulated systems}

The simulated systems consist of one GB1m3 molecule and 
eight crowder molecules, enclosed in a periodic
box with side length 95\,\AA. The eight crowder molecules
are copies of a single protein, either BPTI or GB1.  This setup 
yields crowder densities of $\sim$100 mg/mL, whereas the 
macromolecule densities in cells can be 
$\sim$300--400\,mg/mL.\cite{Vendeville:11} The volume
fraction occupied by the crowders is around 7\%. 
The simulation temperature is set to 332\,K, which is near the 
melting temperature of the free GB1m3 
peptide.\cite{Fesinmeyer:04} For reference, simulations of the 
free peptide are also performed, using the same temperature. 

The GB1m3 peptide is an optimized variant of the second 
$\beta$-hairpin (residues 41--56) in protein GB1, with 
enhanced stability.\cite{Fesinmeyer:04} It differs from 
the original sequence at 7 of 16 positions. To our knowledge,
no experimental structure is available for GB1m3, but its
native fold is expected to be similar to the parent $\beta$-hairpin
in GB1. 

\subsection{Biophysical model}

Our simulations are based on an all-atom protein
representation with torsional degrees of freedom,
and an implicit solvent force field,\cite{Irback:09}
A detailed description of the interaction
potential can be found elsewhere.\cite{Irback:09} In brief, the potential consists
of four main terms, $E = \Eloc +\Eev + \Ehb + \Esc$.  One term ($\Eloc$)
represents local interactions between atoms separated by only a few
covalent bonds. The other, non-local terms
represent excluded-volume effects ($\Eev$), hydrogen bonding ($\Ehb$),
and residue-specific interactions between pairs of side-chains,
based on hydrophobicity and charge ($\Esc$).
In multi-chain simulations, intermolecular interaction terms have the
same form and strength as the corresponding intramolecular ones.
The potential is an effective
energy function, parameterized through folding thermodynamics
studies for a structurally diverse set of peptides and
small proteins.\cite{Irback:09}
Previous applications of the model include folding/unfolding studies
of several proteins with $>$90 residues.\cite{Mitternacht:09,
Jonsson:12,Mohanty:13,Bille:13,Jonsson:13,Petrlova:14} 
Recently, it was used by us to simulate the peptides trp-cage and 
GB1m3 in the presence of protein crowders.\cite{Bille:15,Bille:16}

Our simulations use the same fully atomistic representation of
both the GB1m3 peptide and the crowder proteins. However,  
because of their high thermal stability,\cite{Moses:83,Gronenborn:91}
the crowder proteins are modeled with a fixed backbone, and thus with
side-chain rotations as their only internal degrees of freedom. The assumed
backbone conformations of BPTI and GB1 are model approximations of the
PDB structures 4PTI and 2GB1, derived by MC with
minimization. The structures were selected for both low energy and
high similarity to the experimental structures. The root-mean-square
deviations (RMSDs) from the experimental structures (calculated over
backbone and \Cb\ atoms) are $\lesssim$1\,\AA.

\subsection{MC simulations}
\label{sec:mc}

The systems are simulated using MC dynamics. The simulations
are done in the canonical rather than some generalized 
ensemble.  Also, only ``small-step'' elementary moves are used,
so that the system cannot artificially jump between free-energy 
minima, without having to climb intervening barriers. 
With these restrictions, the simulations
should capture some basics of the long-time dynamics.\cite{Tiana:07} 
Despite the restrictions, the methods are sufficiently fast to
permit the study of the folding and binding thermodynamics of the peptide, 
through simulations containing multiple folding/unfolding and binding/unbinding
events. 

Our move set consists of four different updates:  
(i) the semi-local Biased Gaussian Steps (BGS) method\cite{Favrin:01} for  
backbone degrees of freedom in the peptide, (ii) simple single-angle
Metropolis updates in side chains, (iii) small rigid-body translations of 
whole chains, and (iv) small rigid-body rotations of whole chains.
The ``time'' unit of the simulations is MC sweeps, where one MC 
sweep consists of one attempted update per degree of freedom.
Specifically, each MC sweep consists of 74 attempted moves in the
crowder-free system, whereas the corresponding numbers are
1208 and 1328 with BPTI and GB1 crowders, respectively.    
Note that the average number of attempted
conformational updates of the peptide
per MC sweep is the same in all three cases. 
In the simulations with crowders, the relative fractions of BGS moves, 
side-chain updates, rigid-body translations and rigid-body rotations 
are approximately 0.02, 0.94, 0.02 and 0.02, respectively.

All simulations are run with the program PROFASI,\cite{profasi}
using both vector and thread parallelization. To gather statistics,
a set of independent runs is generated for each system. The number
of runs is 16 with BPTI crowders, 62 with GB1 crowders, and 30 
for the isolated peptide. Each run comprises $40\times10^6$
MC sweeps if crowders are present, and $10\times10^6$ 
MC sweeps without crowders. Compared to the longest relaxation 
times in the respective systems (see below), the individual runs are 
a factor $>$20 longer. 

Several different properties are recorded during the simulations. 
As a measure of the nativeness of the peptide, the number of native H bonds 
present, $\nhb$, is computed, assuming that the native H bonds are 
the same as in the full GB1 protein (PDB code 2GB1). The interaction
of the peptide with surrounding crowder molecules is studied by monitoring
intermolecular H bonds and \Ca-\Ca\ contacts. 
A residue pair is said to be in contact if their \Ca\ atoms are within 8 \AA.    

As input for our TICA and MSM analyses (see below), two sets of parameters 
are stored at regular intervals during the course of the simulations. 
The first set consists of all (non-constant) intramolecular \Ca-\Ca\ distances 
within the peptide, called $r_{ij}$. The second set consists of
intermolecular distances between the peptide and the crowders, called $d_{ij}$. 
Specifically, $d_{ij}$ denotes the shortest \Ca-\Ca\ distance between 
peptide residue $i$ and residue $j$ in any of the crowder molecules.   

\subsection{TICA and MSM analysis}
\label{sec:mm}

TICA can be used as a dimensionality reduction method.
It is somewhat similar to principal component analysis,  but identifies 
high-autocorrelation (or slow) rather than high-variance coordinates. 
Given time trajectories of
a set of parameters, $\{o_n\}$ (in our case, 
the distances $r_{ij}$ and $d_{ij}$, see above),   
one constructs the time-lagged covariance matrix
$c_{nm}(\tcm)=\ev{o_n(t)o_m(t+\tcm)}_t-\ev{o_n(t)}_t\ev{o_m(t+\tcm)}_t$,
where $\tcm$ is the lag time and $\ev{\cdot}_t$ denotes an average 
over time $t$. By solving the (generalized) eigenvalue problem
$\textbf{C}(\tcm)\hat{\textbf{v}}_i=\lcm_i\textbf{C}(0)\hat{\textbf{v}}_i$,
slow linear combinations of the original parameters can 
be identified. A more advanced method for identifying 
slow modes is to construct MSMs.

To build an MSM, the state space needs to be 
discretized. In our calculations, following 
Ref.~\citen{PerezHernandez:13}, the discretization is achieved by  
clustering the data with the $k$-means algorithm\cite{Lloyd:82} 
in a low-dimensional subspace spanned by slow TICA 
coordinates.  By computing the probabilities of transition 
among these clusters in a time $\ttm$ (which, like $\tcm$, 
is an adjustable parameter), a transition matrix is obtained. 
Assuming Markovian dynamics, the eigenvectors of this matrix 
have relaxation times given by 
\begin{equation}
\tilde t_i=-\ttm/\ln\ltm_i(\ttm)
\label{eq:timescales}
\end{equation}
where $1=\ltm_0>\ltm_1\ge\ltm_2\ge...>0$ are the eigenvalues. 
The eigenvalue $\ltm_0$ corresponds 
to a stationary distribution ($\tilde t_0=\infty$), 
whereas all other eigenvalues correspond to relaxation modes with 
finite timescales $\tilde t_i$. The timescales obtained using  Eq.~\ref{eq:timescales} 
are expected to reproduce the dominant relaxation times of the full system 
if  the discretization is sufficiently fine,\cite{Kube:07,Djurdjevac:12} or 
if the lag time is sufficently large.\cite{Djurdjevac:12,Prinz:14}     

There are several software packages available for TICA and MSM 
analysis.\cite{pyemma,wordom,metagui,msmbuilder} 
Our calculations are done using the 
pyEMMA software.\cite{pyemma}  

\subsection{Timescales from autocorrelations of MSM eigenfunctions}
\label{sec:aa}

Another way of estimating relaxation times from an MSM is  
to compute autocorrelations of the eigenfunctions.    
The (normalized) autocorrelation function of a general property $f$ is given by    
$C_f(\tau)=[\ev{f(t)f(t+\tau)}_t-\ev{f(t)}_t\ev{f(t+\tau)}_t]/\sigma_f^2$,
where $\sigma_f^2$ is the variance of $f$.    
Let $\psi_i^{\text{MSM}}$ be the $i$th eigenfunction of 
a given MSM, and let $\psi_i$ be the true $i$th eigenfunction 
of the system's time transfer operator.\cite{Prinz:11} The autocorrelation function of 
$\psi_i^{\text{MSM}}$, $C_i(\tau)$, 
may be expanded as
\begin{equation}
C_i(\tau)=\sum_j c_j\text{e}^{-\tau/t_j}
\label{eq:expand}
\end{equation}
where $t_j$ is the exact $j$th relaxation time. The coefficients $c_j$ are
given by $c_j=|\ev{\psi_j,\psi_i^\text{MSM}}|^2$, where the overlap 
$\ev{\psi_j,\psi_i^\text{MSM}}$ can be expressed 
as an average with respect to the stationary distribution, $\mu(x)$: 
$\ev{\psi_j,\psi_i^\text{MSM}}=\int dx\mu(x)\psi_j(x)\psi_i^\text{MSM}(x)$.
Note that $\psi_j$ and $\psi_i^\text{MSM}$ have mean zero and 
unit norm. In Sect.~\ref{sec:results}, overlaps between pairs of general 
functions are computed in the same way, after shifting and 
normalizing the functions to zero mean and unit norm.  

Now, if $\psi_i^{\text{MSM}}$
is a good approximation of $\psi_i$, then $c_j\ll c_i$ for $j\ne i$.  
If this holds, $C_i(\tau)$ decays approximately as $\text{e}^{-\tau/t_i}$  
for not too large $\tau$ (compared to $t_i$), so that $t_i$ 
can be estimated through a simple exponential fit.     

The calculations discussed below  
use data for $C_i(\tau)$ in the range of $\tau$ where 
$0.2<C_i(\tau)<0.8$. Over this range, $C_i(\tau)$ is to
a good approximation single exponential for all MSM eigenfunctions 
studied. The upper bound on $\tau$ is set primarily by statistical 
uncertainties, rather than by deviations from single-exponential 
behavior. 

\section{Results}
\label{sec:results}

Our analysis of the GB1m3 peptide in the three simulated 
systems (with BPTI crowders, with GB1 crowders, without crowders)
can be divided into two parts. First, equilibrium free-energy surfaces
are constructed, using TICA coordinates. Second, the dynamics  
are investigated, using MSM techniques. 

\subsection{Free-energy landscapes}

\begin{figure}[t]
\centering
\includegraphics[width=8cm]{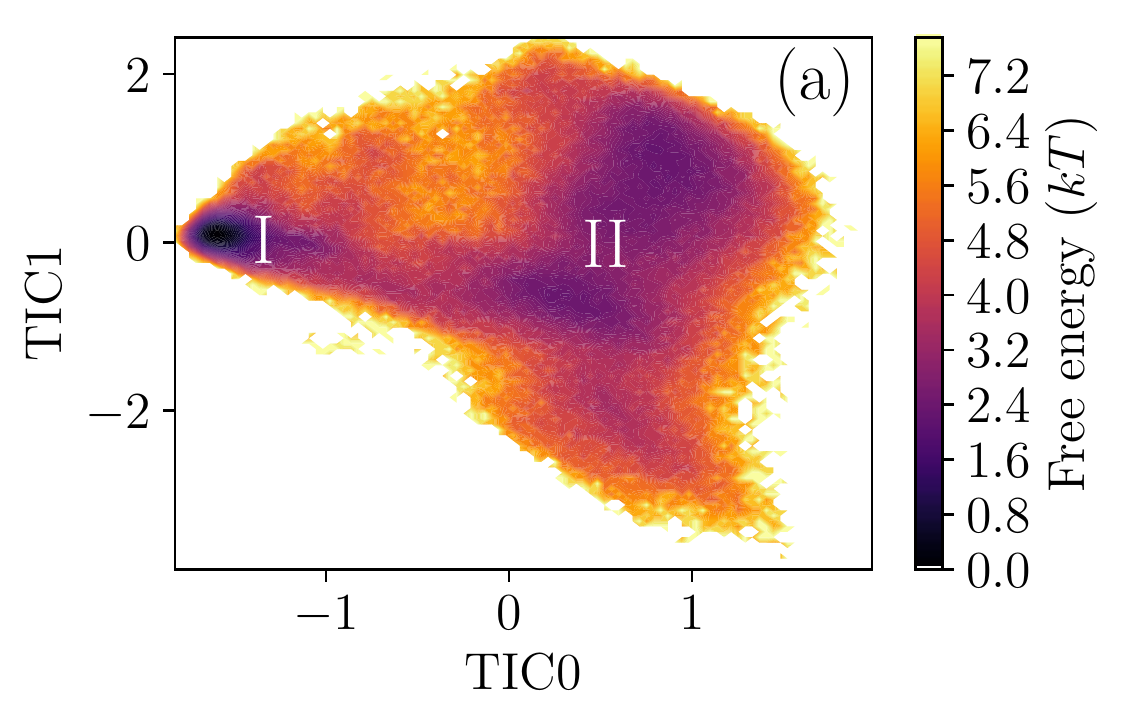}
\includegraphics[width=8cm]{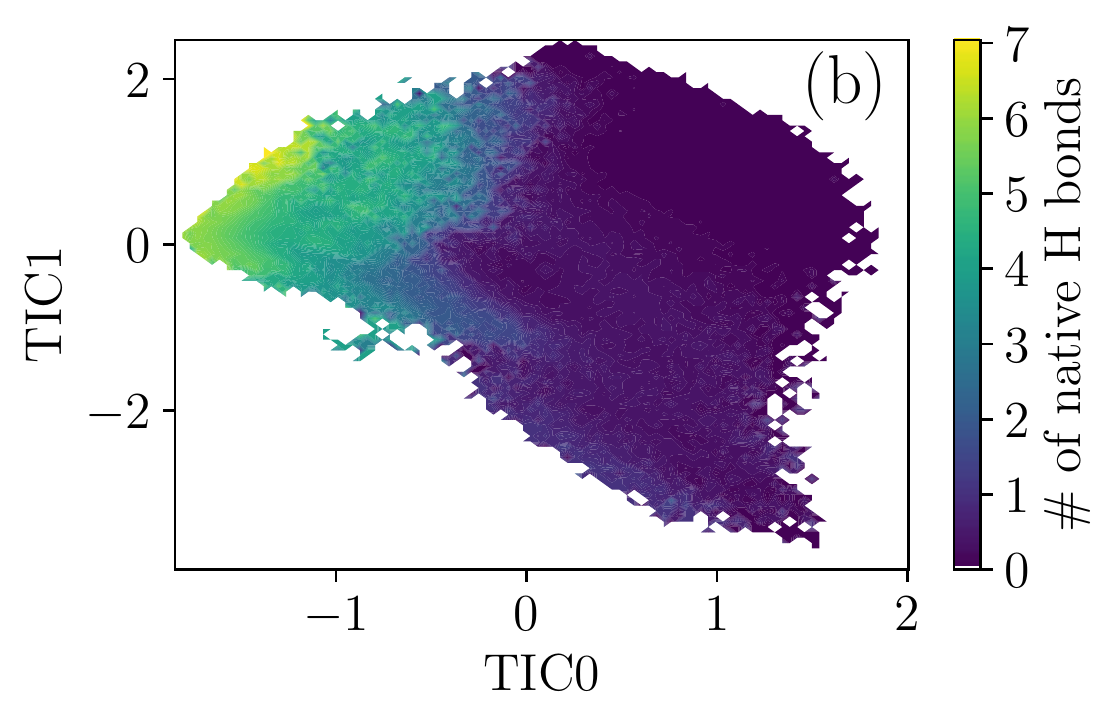}
\caption{
(a) Free energy of the isolated GB1m3 peptide, calculated as a function of 
the two slowest TICA coordinates, TIC0 and TIC1. Major minima
are labeled by Roman numerals.
(b) The dependence of the number of native H bonds, $\nhb$, on these coordinates.
Here, each stored conformation is represented by a point in the 
TIC0,TIC1-plane, 
in a color determined by the value of $\nhb$. Smoothing is 
applied to improve readability. The TICA lag time  
is set to $\tcm=10^3$\,MC sweeps. 
\label{fig:modes_free}}
\end{figure}

It is instructive to begin with the isolated GB1m3 peptide, whose 
folding thermodynamics have been studied before using the same 
model.\cite{Irback:09} This study found that the isolated peptide
folds in a cooperative manner, and that the number of native H bonds
present, $\nhb$, is a useful folding coordinate that has a bimodal 
distribution at the melting temperature. 
Figure~\ref{fig:modes_free}a shows the free energy 
of the isolated GB1m3, calculated as a function of the
two slowest TICA coordinates, TIC0 and TIC1. The free-energy 
surface exhibits two major minima, labeled I and II, 
which are well separated in the TIC0 direction. From Fig.~\ref{fig:modes_free}(b),
it can be seen this coordinate is strongly (anti-) correlated with $\nhb$.  
This correlation implies that the peptide is native-like in 
free-energy minimum I, and unfolded in minimum II.   
 
\begin{figure}[t]
\centering
\includegraphics[width=8cm]{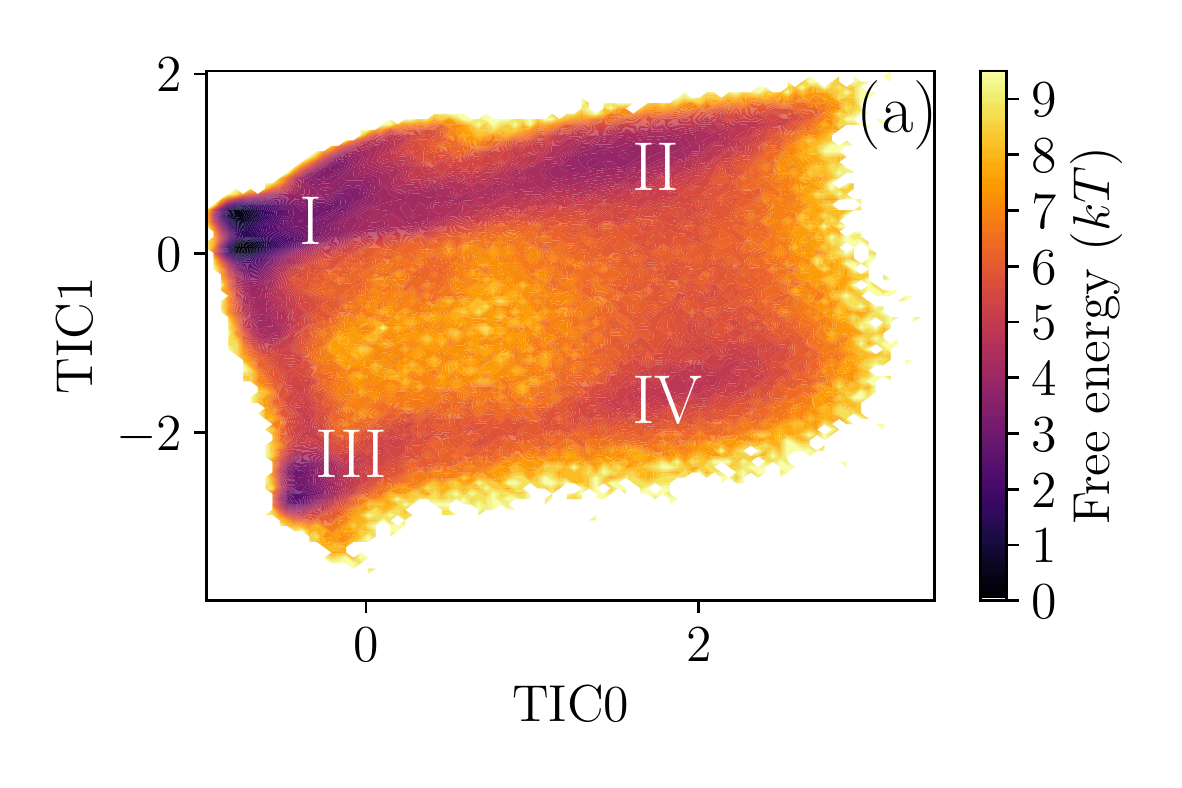}\\

\vspace{-6pt}

\includegraphics[width=8cm]{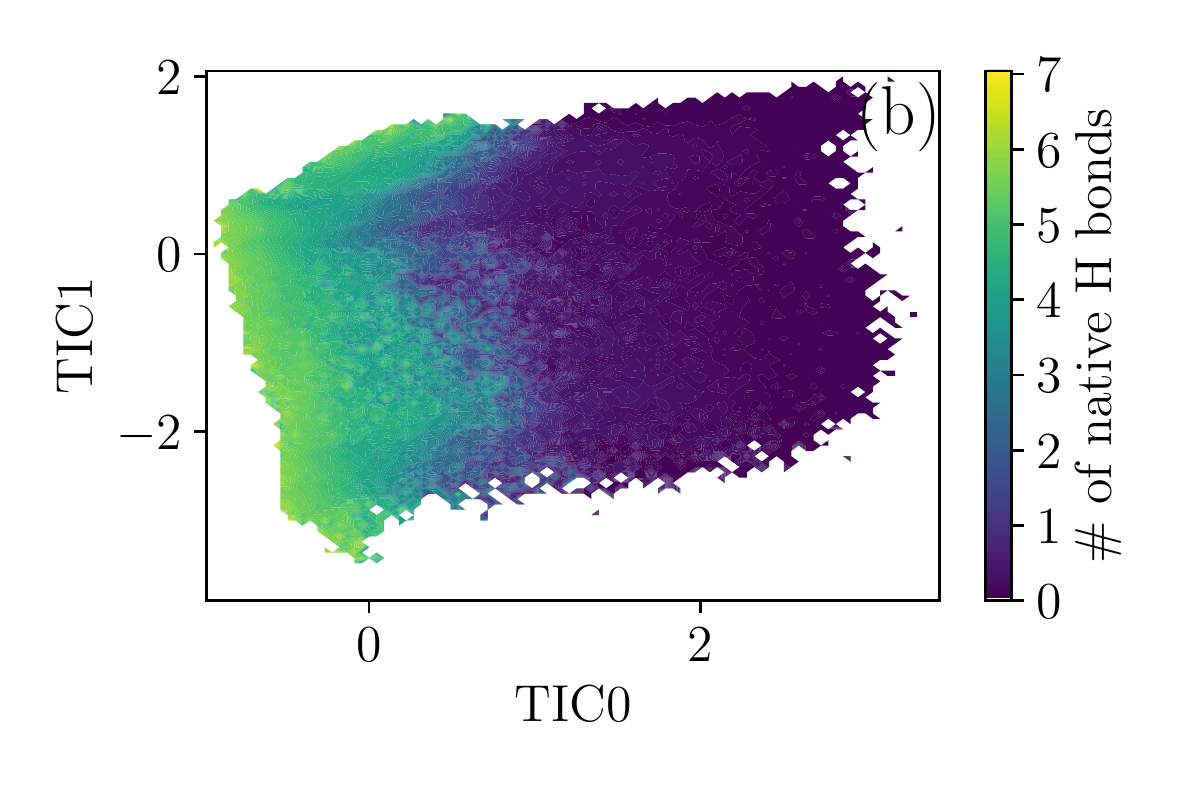}
\includegraphics[width=8cm]{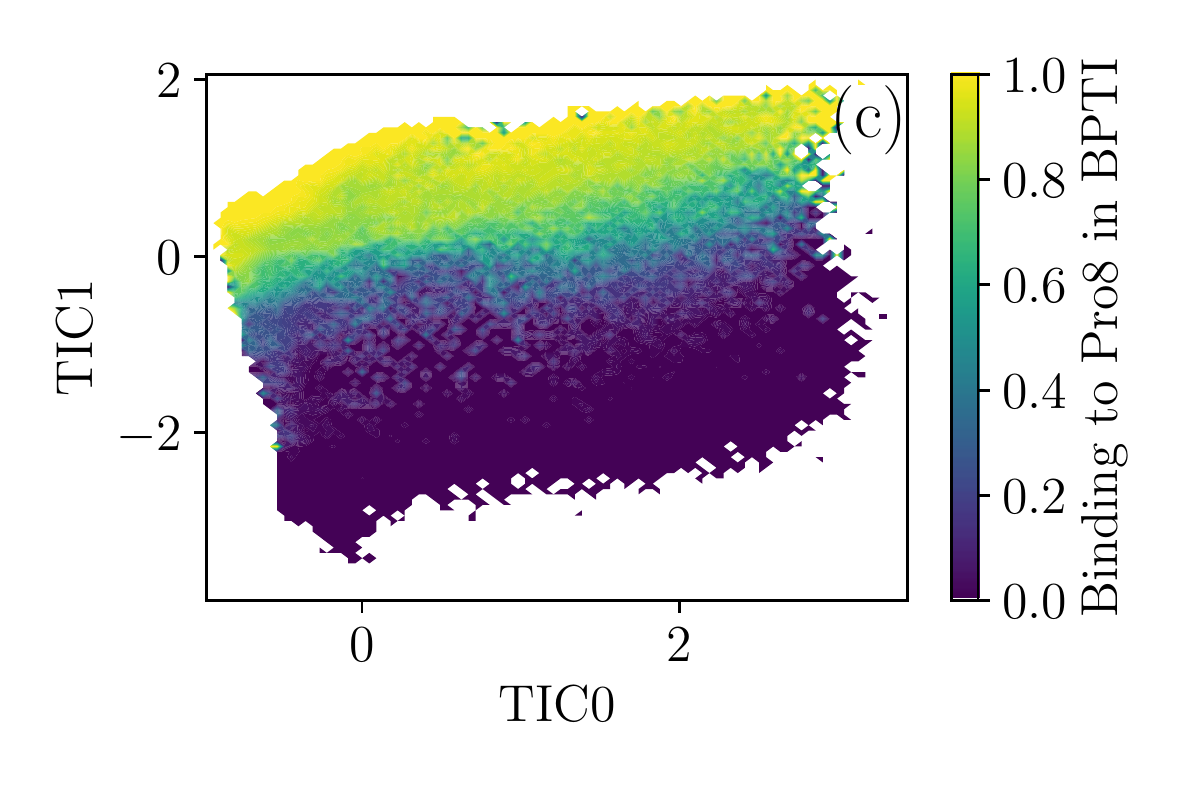}
\caption{
Characterization of the GB1m3 peptide in the presence of BPTI crowders, using
the two slowest TICA coordinates, TIC0 and TIC1. (a) Free energy. Major minima
are labeled by Roman numerals.
(b) The number of native H bonds present in the peptide, $\nhb$.  
(c) A function which is unity whenever there is at least one residue-pair 
\Ca-\Ca\ contact between the peptide and a Pro8 BPTI residue,
and zero otherwise (drawn using smoothing).    
The contact cutoff distance is 8 \AA.
The TICA lag time is set to $\tcm=10^3$\,MC sweeps.  
\label{fig:modes_BPTI}}
\end{figure}

We now turn to the system where GB1m3 is surrounded by BPTI crowders. 
Here, the TICA coordinates are linear combinations of both intra-
and intermolecular distances ($r_{ij}$ and $d_{ij}$; see Sec.~\ref{sec:mc}). Calculated 
as a function of the two slowest TICA coordinates, the free energy exhibits   
four major minima, labeled I--IV~(Fig.~\ref{fig:modes_BPTI}(a)).  
To characterize the minima, an interpretation of the  
TIC0 and TIC1 coordinates is needed.
As in the previous case, TIC0 
is strongly correlated with $\nhb$~(Fig.~\ref{fig:modes_BPTI}(b)), and 
thus linked to the degree of nativeness. 
Inspection of the eigenvector corresponding to TIC1
suggests that this coordinate depends strongly on certain
peptide-crowder distances $d_{ij}$ 
involving the BPTI residue Pro8, which is part of a sticky patch 
on the BPTI surface.\cite{Bille:16} Motivated by this observation, 
Fig.~\ref{fig:modes_BPTI}(c) shows the TIC0,TIC1-dependence
of a function defined to be
unity whenever there is at least one residue-pair contact 
between the peptide and a Pro8 BPTI residue, and zero otherwise
(smoothing is used). This function is indeed strongly correlated with 
TIC1. Therefore, the main free-energy minima can be 
classified based on whether or not the peptide is native-like, and 
whether or not the peptide forms any Pro8 BPTI contact.             
The peptide is native-like and bound in minimum I, which 
actually can be split into two distinct subminima, corresponding 
to two preferred orientations of the folded and bound peptide. 
In the remaining three main minima,  
the peptide is either unfolded and 
bound (minimum II), native-like and unbound (minimum III),  or
unfolded and unbound (minimum IV).

\begin{figure}[t!]
\centering
\includegraphics[width=8cm]{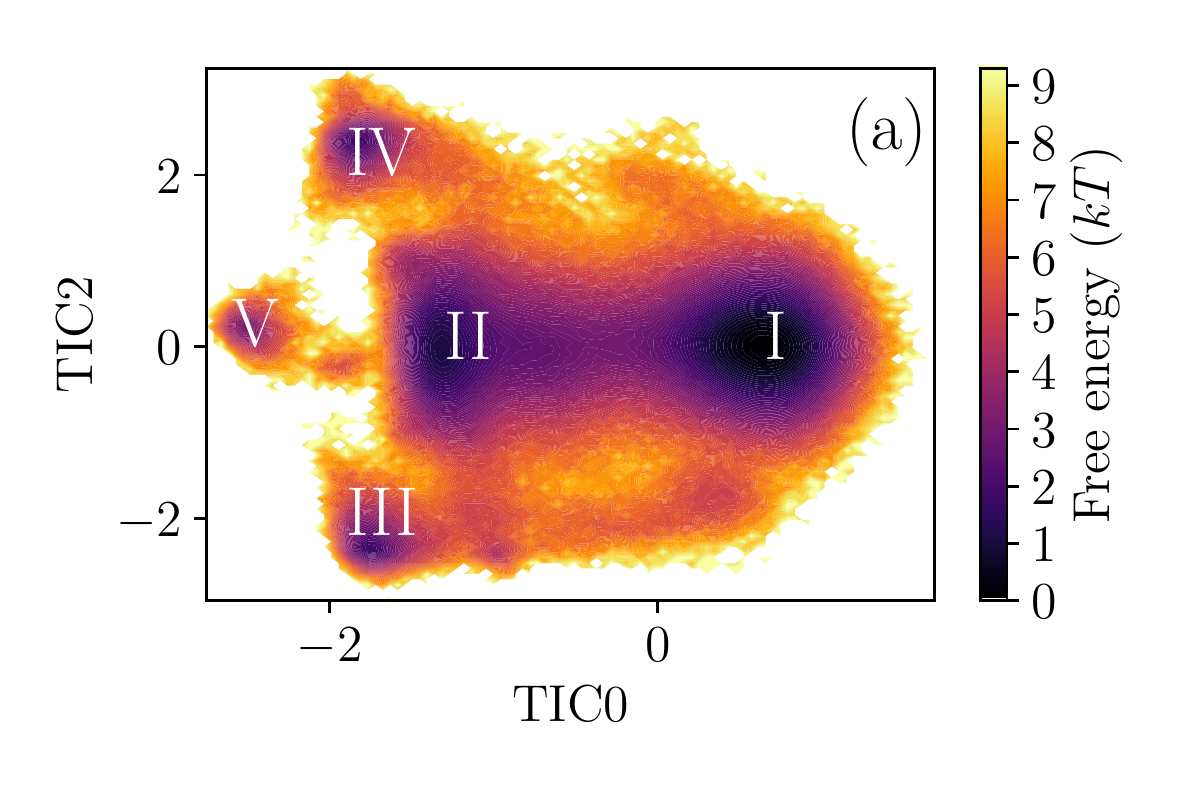}
\includegraphics[width=8cm]{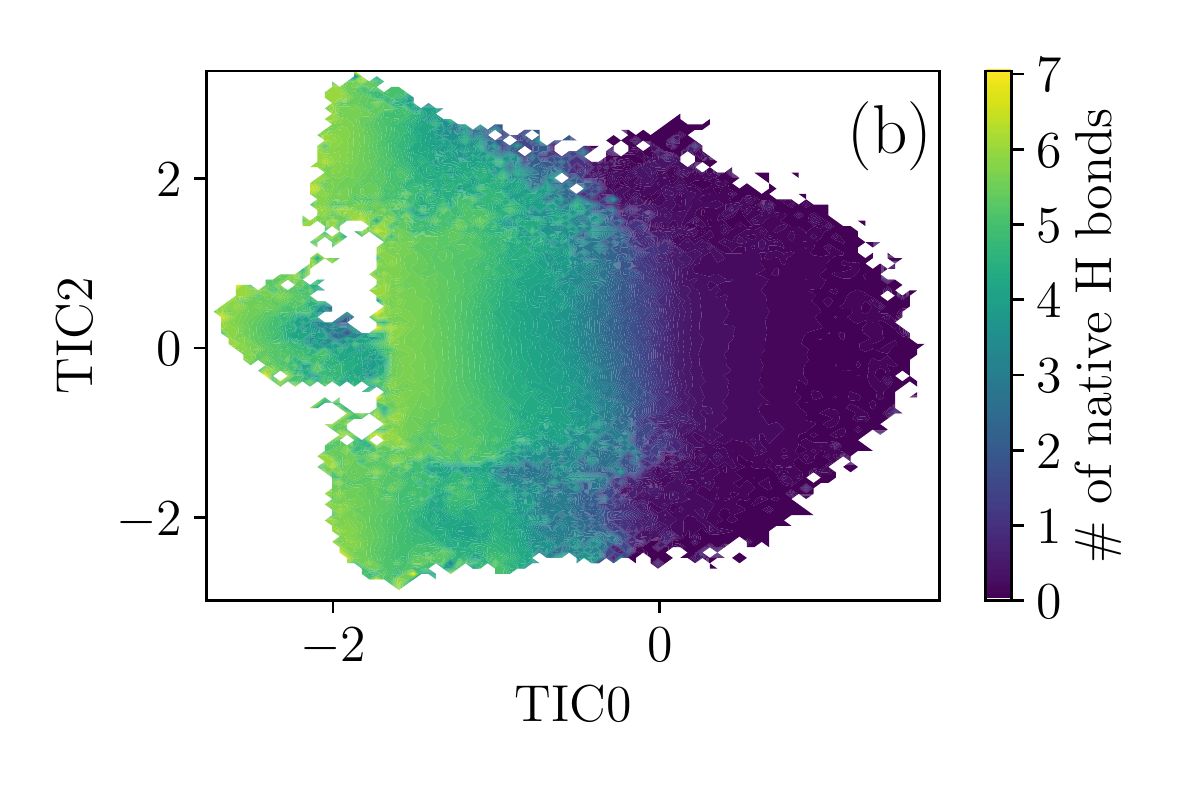}

\vspace{-6pt}

\includegraphics[width=8cm]{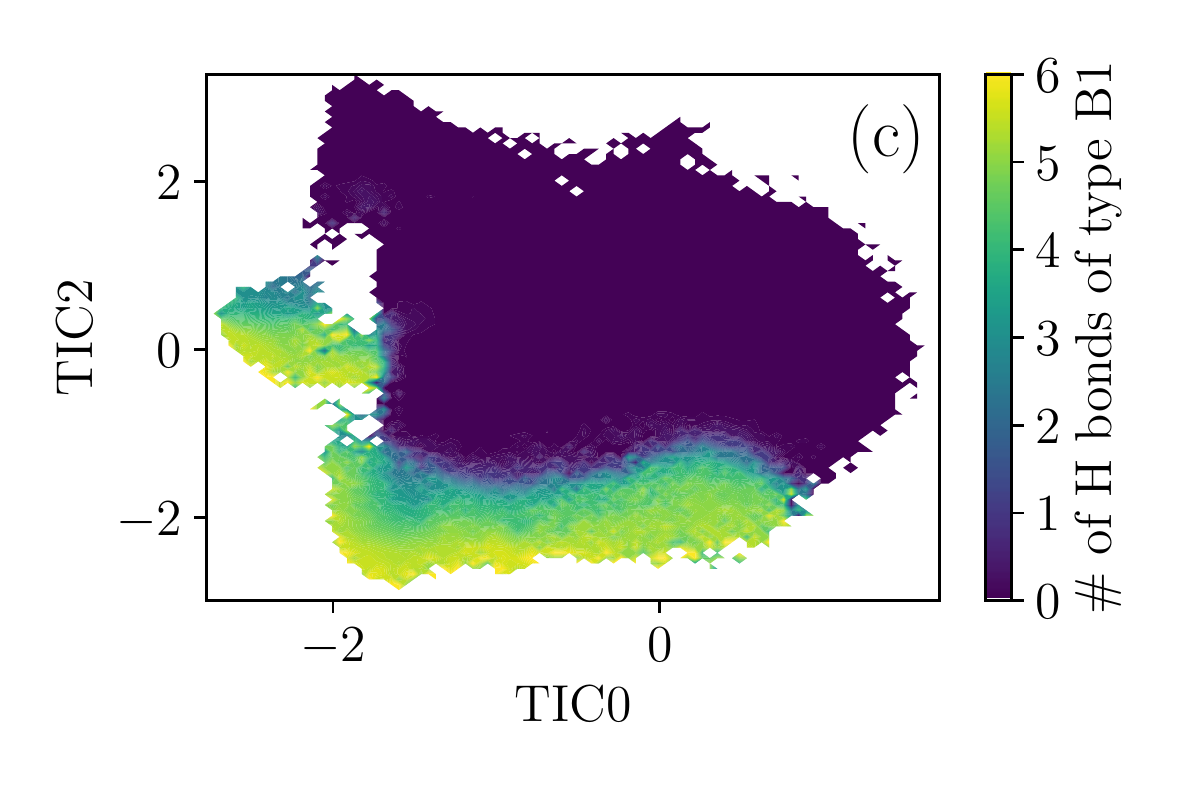}
\includegraphics[width=8cm]{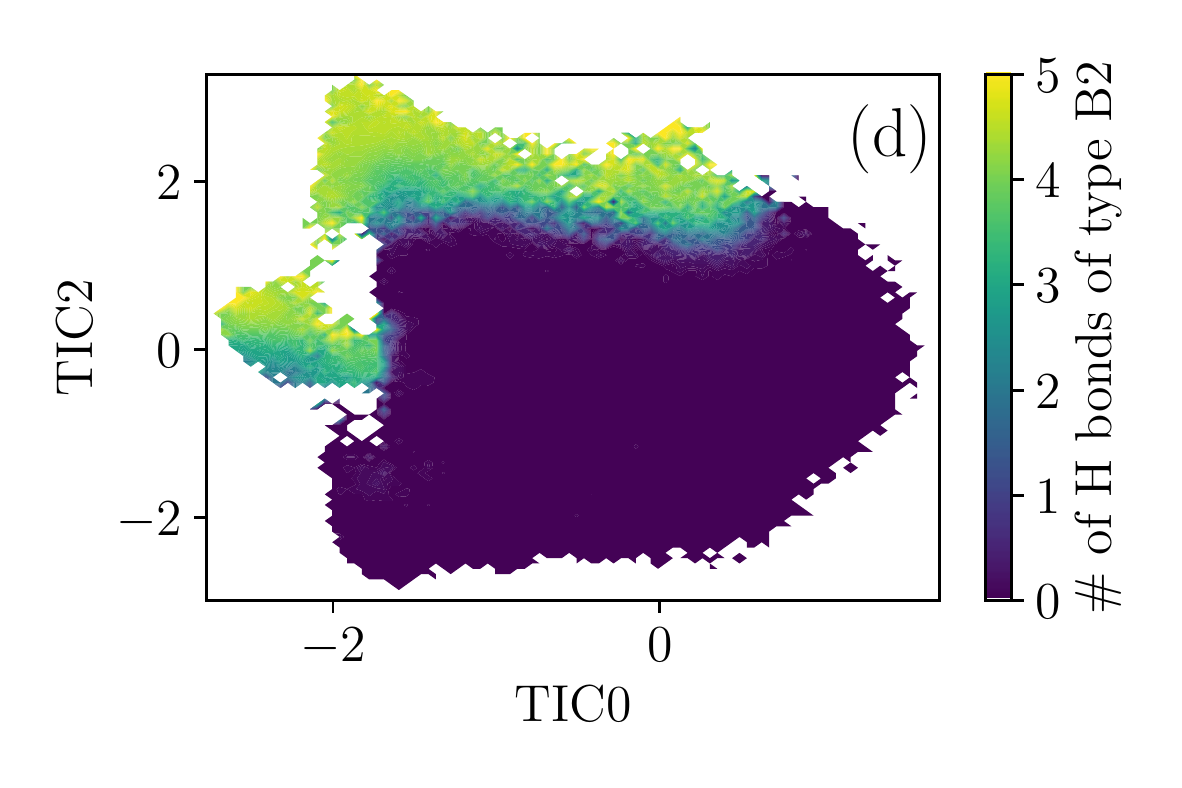}
\caption{
Characterization of the GB1m3 peptide in the presence of GB1 crowders, using
the slowest and third-slowest TICA coordinates, TIC0 and TIC2. (a) Free energy. Major minima
are labeled by Roman numerals.
(b) The number of native H bonds present in the peptide, $\nhb$ 
(c,d) The numbers of present H bonds associated with the peptide-crowder binding modes 
B1 and B2 (see the supplementary material, Fig.~S2), respectively. 
The TICA lag time is set to $\tcm=20\times10^3$\,MC sweeps. 
\label{fig:modes_GB1}}
\end{figure}

With GB1 crowders, the free energy of GB1m3 exhibits five 
well-separated and easily visible minima (Fig.~\ref{fig:modes_GB1}(a)) 
when calculated as a function of the slowest and third-slowest TICA coordinates.
The TIC0,TIC2-plane is used here because two of the minima (III and IV) 
cannot be distinguished in the TIC0,TIC1-plane (see the supplementary material, Fig.~S1). 
The TIC0 coordinate is again correlated with the degree of 
nativeness of the peptide (Fig.~\ref{fig:modes_GB1}(b)). Proper interpretation of  
the TIC2 coordinate requires knowledge of the preferred peptide-crowder
binding modes. It turns out that there are two preferred binding modes, 
called B1 and B2.  In both cases, binding occurs through $\beta$-sheet extension; 
the edge strand $\beta$3 (residues 42--46) of GB1 binds to either the 
first or the second strand of the folded GB1m3 $\beta$-hairpin. The binding modes can be described in 
terms of the H bonds involved (see the supplementary material, Fig.~S2). 
Figures~\ref{fig:modes_GB1}(c) and \ref{fig:modes_GB1}(d) show how 
the presence of H bonds associated with the respective modes vary with
TIC0 and TIC2. Apparently, low and high TIC2 signal B1 and B2 binding, 
respectively. A similar analysis of TIC1 shows that this coordinate separates 
bound and unbound states, but discriminates poorly between the B1 and B2
modes (see the supplementary material, Fig.~S1(c,d)). 
The isolated island at low TIC0 and intermediate TIC2 
stems from simultaneous binding of the peptide via both modes, to 
two crowder molecules. Based on the above observations, the free-energy minima 
in Fig.~\ref{fig:modes_GB1}(a) can be described as follows. In minima 
I and II, the peptide is unfolded and native-like, respectively, and neither 
B1 nor B2 binding occurs. In the remaining three minima, the peptide is native-like 
and bound. The mode of binding is either B1 (minimum III), B2 (minimum IV) 
or both (minimum V).      

It is worth noting that the interpretation of the TIC0 coordinates of the 
BPTI and GB1 systems is not necessarily the same, although TIC0 is a strongly 
correlated with folding in both cases. In the GB1 system, TIC0 is correlated not 
only with folding but also with double binding (Fig.~\ref{fig:modes_GB1}(c,d)). 
By contrast, in the BPTI system, the correlation between TIC0 and 
the binding coordinate is weak (Fig.~\ref{fig:modes_BPTI}(c)). 

To sum up, the results of this section show that TICA provides useful 
coordinates for describing the free energy of the peptide in the different systems. 
Using a few slow TICA coordinates, the main free-energy minima
can be identified.

\subsection{Dynamics}

TICA provides a first approximation of the slow modes. For a more 
detailed investigation of the dynamics of the peptide in our 
simulations with crowders, MSMs are constructed as described in 
Sec.~\ref{sec:mm}, for a range of lag times $\ttm$. Relaxation times are 
estimated by two methods: (i) from MSM eigenvalues 
(Eq.~\ref{eq:timescales}), and (ii) by fits to autocorrelation 
data for MSM eigenfunctions (Sec.~\ref{sec:aa}). Illustrations of 
how the main MSM eigenfunctions are related to the 
TICA modes discussed above can be found in the
supplementary material (Figs. S3--S6).   

Figure~\ref{fig:timescales_vs_ttm_BPTI}(a) shows estimates of the four
longest relaxation times in the system with BPTI crowders, 
as obtained by the above-mentioned methods. 
As expected, the eigenvalue-based estimates have systematic errors for small lag 
times $\ttm$. To keep this error low, $\ttm$ has to be comparable to the timescale 
in question. The estimates based on autocorrelation analysis 
depend, by contrast, only very weakly on $\ttm$. This behavior suggests 
that the true relaxation times can be estimated from the MSM 
eigenfunctions even if $\ttm$ is relatively small. Consistent with 
this, a further test shows that the shape of the slowest MSM eigenfunction 
depends only weakly on $\ttm$. Here, pairwise overlaps (see Sect.~\ref{sec:aa})
were computed between variants of this function obtained for different $\ttm$. 
The overlap was $\ge$0.96 for all possible pairs of $\ttm$.

\begin{figure}[t]
\centering
\includegraphics[width=8cm]{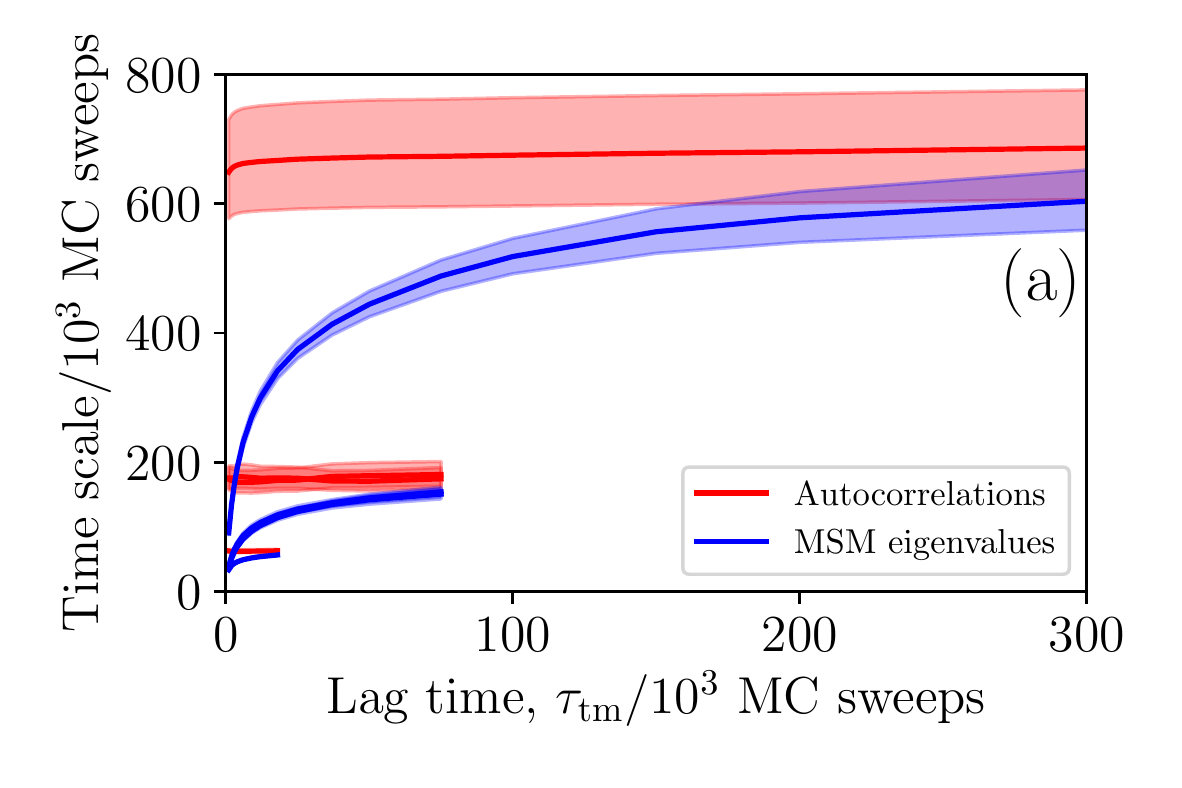}
\includegraphics[width=8cm]{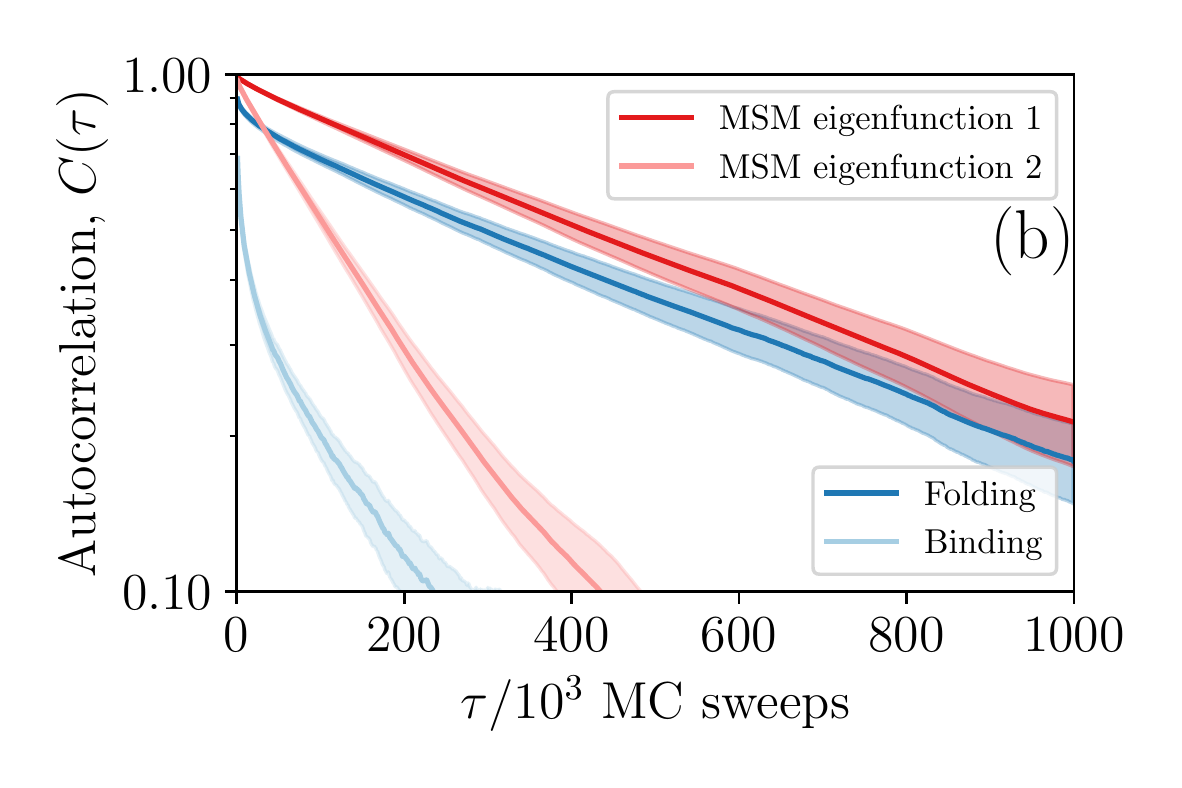}
\caption{
Long-time dynamics of GB1m3 in the presence of 
BPTI crowders. 
Shaded areas indicate statistical 1$\sigma$ errors.
(a) Estimates of the four 
longest relaxation times, as obtained using  
MSM eigenvalues (Eq.~\ref{eq:timescales}; blue curves) and 
autocorrelation analysis (Sec.~\ref{sec:aa}; red curves).
The data are plotted against the lag time $\ttm$ of the MSM 
transition matrix. The second and third longest timescales 
are very similar. In building the MSMs, data were clustered 
in the space spanned by the four slowest TICA modes
(using $\tcm=10^3$ MC sweeps), into 800 clusters.
(b) Autocorrelation functions, $C(\tau)$, 
for the two slowest MSM eigenfunctions ($\ttm=25\times 10^3$ MC sweeps), 
the folding variable $\nhb$ (Fig.~\ref{fig:modes_BPTI}(b)), and 
the binding variable studied in Fig.~\ref{fig:modes_BPTI}(c).
\label{fig:timescales_vs_ttm_BPTI}}
\end{figure}

Figure~\ref{fig:timescales_vs_ttm_BPTI}(b) compares the
raw autocorrelation functions for the two slowest MSM 
eigenfunctions to those for the folding and binding 
coordinates studied in Figs.~\ref{fig:modes_BPTI}(b)
and \ref{fig:modes_BPTI}(c), 
respectively. One observation 
that can be made is that the autocorrelations
of the folding and binding coordinates, not unexpectedly, 
show clear deviations from single-exponential behavior 
at small $\tau$. The MSM eigenfunctions are, as intended 
by construction, much closer to single exponential,
which facilitates the extraction of relaxation times. 

Another observation from Fig.~\ref{fig:timescales_vs_ttm_BPTI}(b) 
is that, except at small $\tau$, the autocorrelations of the first MSM 
eigenfunction and the folding coordinate decay at very similar rates.
A close relationship between these two functions is indeed suggested 
from a comparison of Figs.~\ref{fig:modes_BPTI}(b)  and 
S4(a) (see the supplementary material). This conclusion is further strengthened
by their overlap (about 0.88). The autocorrelation function for the 
second MSM eigenfunction somewhat resembles that for the 
binding coordinate (Fig.~\ref{fig:timescales_vs_ttm_BPTI}(b)), 
but the overlap is not very large (about 0.36); the binding coordinate 
overlaps significantly with other MSM eigenfunctions as well. 
Thus, while the second eigenfunction probably is related to binding, 
that relationship is not fully captured by the binding coordinate. 

\begin{figure}[t]
\centering
\includegraphics[width=8cm]{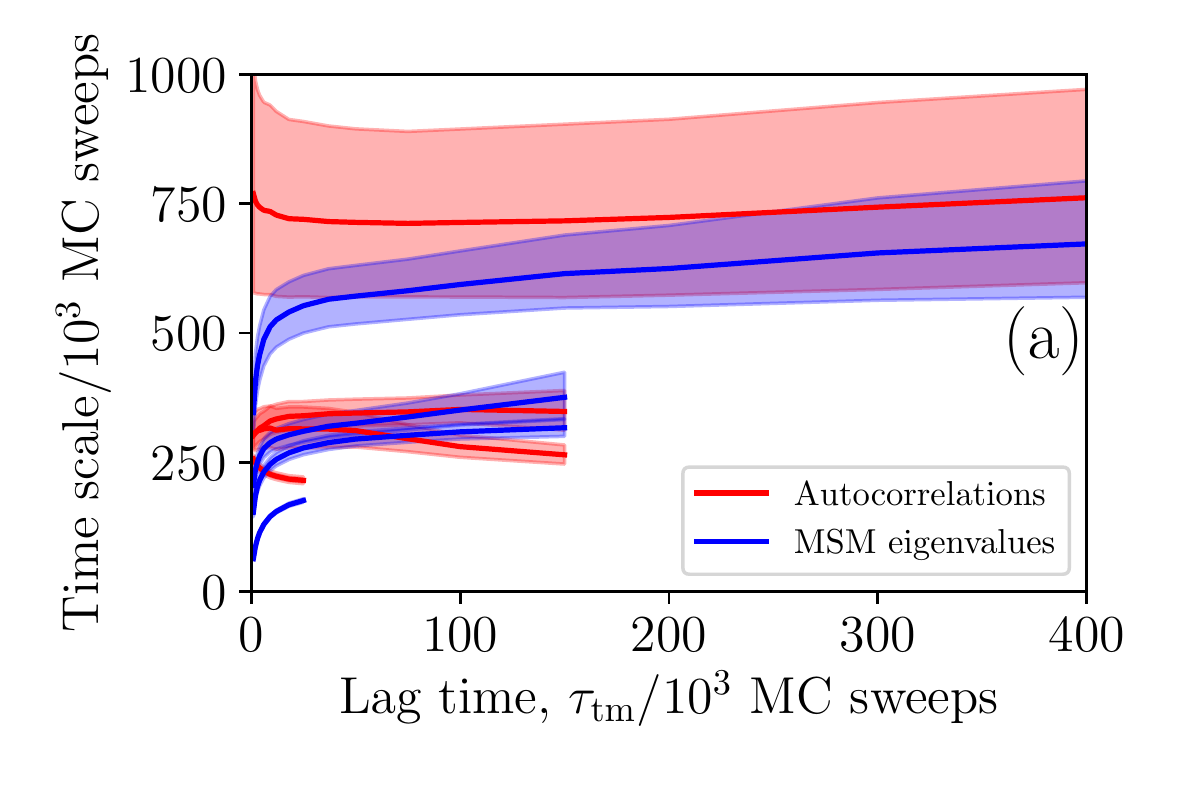}
\includegraphics[width=8cm]{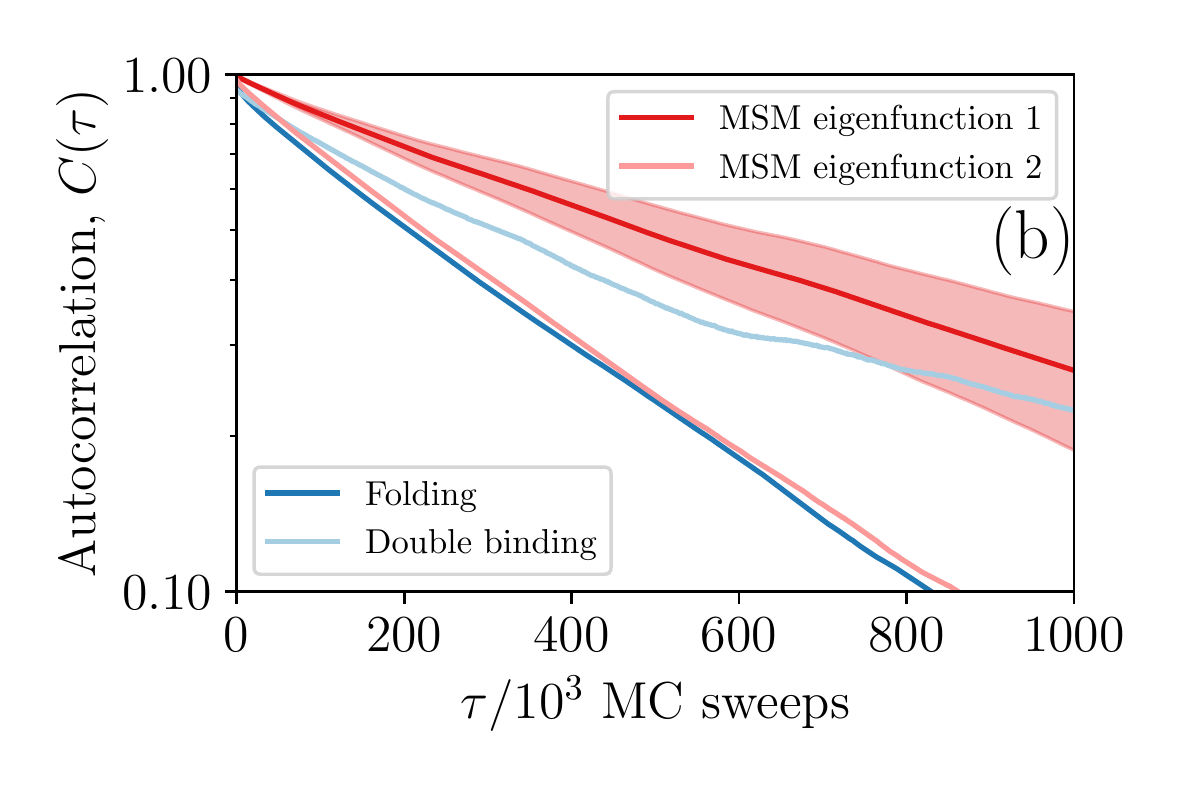}
\caption{
Long-time dynamics of GB1m3 in the presence of GB1 crowders. 
Shaded areas indicate statistical 1$\sigma$ errors.
(a) Estimates of the four 
longest relaxation times, as obtained using
MSM eigenvalues (Eq.~\ref{eq:timescales}; blue curves) and 
autocorrelation analysis (Sec.~\ref{sec:aa}; red curves).
The data are plotted against the lag time $\ttm$ of the MSM 
transition matrix. In building the MSMs, data were clustered 
in the space spanned by the three slowest TICA modes
(using $\tcm=20\times 10^3$ MC sweeps), into 1574 clusters.
(b) Autocorrelation functions, $C(\tau)$, 
for the two slowest MSM eigenfunctions ($\ttm=25\times 10^3$ MC sweeps), 
the folding variable $\nhb$ (Fig.~\ref{fig:modes_GB1}(b)),
and the binding variable $\chi_\text{b}$~(see text). For clarity, statistical 
errors are shown only for one of the four functions. The 
statistical uncertainties are somewhat larger for the binding variable 
$\chi_\text{b}$ than they are for the other three functions.  
\label{fig:timescales_vs_ttm_GB1}}
\end{figure}

Figure~\ref{fig:timescales_vs_ttm_GB1} shows data from our 
simulations with GB1 crowders. 
The statistical uncertainties are larger for this system. The main reason 
for this is that transitions to and from free-energy minimum V 
(Fig.~\ref{fig:modes_GB1}(a)), where the peptide simultaneously binds two crowder 
molecules, occur only rarely in the simulations. Nevertheless, after increasing
the number of runs from 16 for the BPTI system to 62, our total data set 
contains about 30 independent visits to this minimum, and some   
clear trends can be seen. The estimated relaxation times 
follow the same pattern as with BPTI crowders; 
the estimates based on MSM eigenvalues converge 
only slowly with increasing $\ttm$, whereas those 
based on autocorrelation analysis are essentially constant  
down to small $\ttm$ (Fig.~\ref{fig:timescales_vs_ttm_GB1}(a)).  
However, in the GB1 system, the first MSM eigenfunction is more 
closely linked to binding than to folding. To show this, a binary 
function sensitive to simultaneous binding of the peptide to two crowder molecules   
is calculated. Specifically, this function is defined as $\chi_\text{b}=\chi_1\chi_2$, 
where $\chi_i$ is unity if at least three of the H bonds associated with binding 
mode $i$ (see the supplementary material, Fig.~S2) are present, and $\chi_i=0$ otherwise.    
Figure~\ref{fig:timescales_vs_ttm_GB1}(b) shows autocorrelation 
data for the two slowest MSM eigenfunctions,
the folding coordinate ($\nhb$), and the function $\chi_\text{b}$. 
The $\nhb$ and $\chi_\text{b}$ functions are natural candidates for the slowest 
modes, since they are both highly correlated with TIC0. It turns out that 
the autocorrelation function of $\chi_\text{b}$ decays slower than that 
of $\nhb$, and at a rate comparable to that for the first MSM 
eigenfunction (Fig.~\ref{fig:timescales_vs_ttm_GB1}(b)).   
Consistent with this, the first MSM eigenfunction has a 
larger overlap with the binding function $\chi_\text{b}$ (about 0.76) 
than it has with the folding coordinate (about 0.44).      
   
\begin{figure}[t]
\centering
\includegraphics[width=8cm]{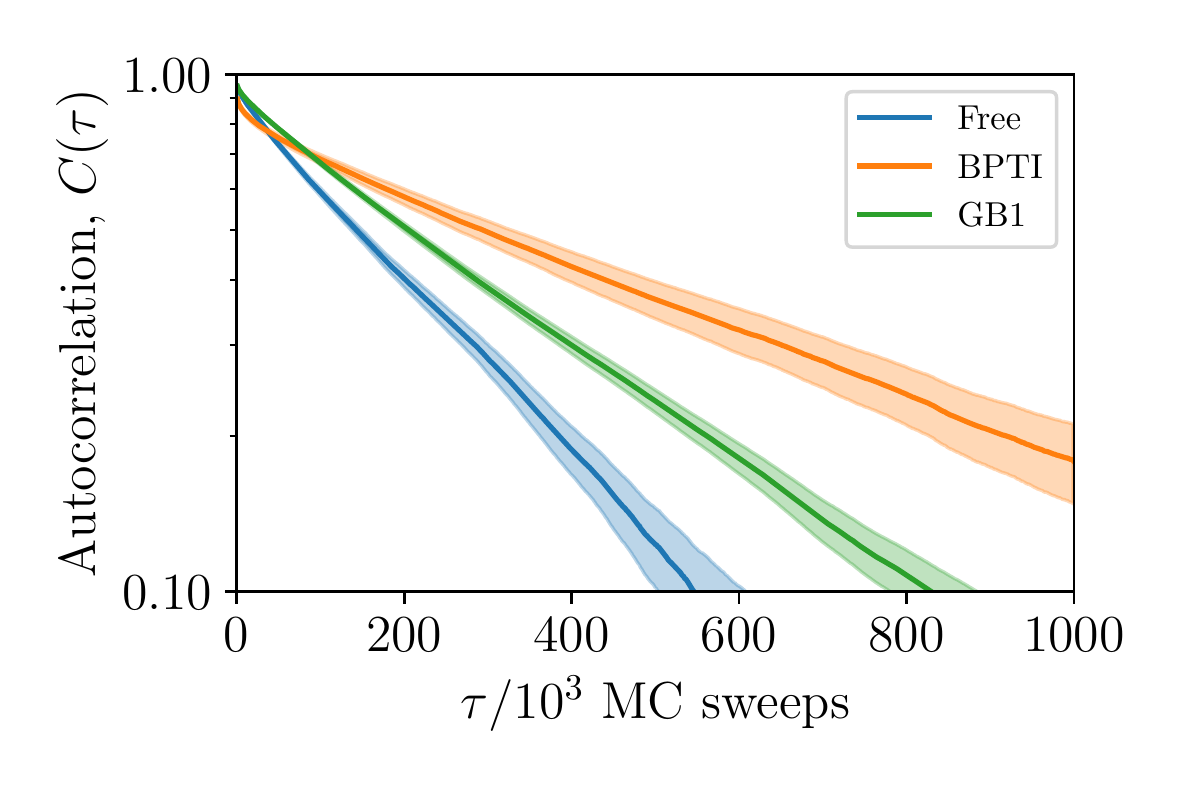}
\caption{
Autocorrelation function, $C(\tau)$, for the 
folding variable $\nhb$ (the number of native H bonds 
present in the peptide), as obtained without crowders, with BPTI 
crowders and with GB1 crowders. Table~\ref{tab:rates} 
shows apparent folding rates $k$ obtained by exponential fits
to the data.  Shaded areas indicate statistical 1$\sigma$ errors.
\label{fig:autocorr_folding}}
\end{figure}

Finally, we compute and compare the folding and unfolding rates of 
the peptide, $\kf$ and $\ku$, in the three simulated environments. 
To this end, we determine the native-state 
probability, $\Pn$ (with the peptide being defined as native if $\nhb\ge 3$),
and the apparent folding/unfolding rate, $k=\kf+\ku$. The rate $k$
is obtained by a fit to autocorrelation data for the folding coordinate 
$\nhb$ (Fig.~\ref{fig:autocorr_folding}). Knowing $k$ and $\Pn$ and 
assuming a simple folded/unfolded two-state behavior, 
$\kf$ and $\ku$ can be computed ($\kf=k\Pn$, $\ku=k-\kf$). 
Our data for $\Pn$, $k$, $\kf$ and $\ku$ are summarized in 
Table~\ref{tab:rates}. The BPTI crowders cause a 
considerable stabilization of the peptide (increased $\Pn$),
and a marked decrease in $k$. The decrease in $k$
can be attributed to a lower $\ku$; no significant change
in $\kf$ is observed. With GB1 crowders, a similar pattern is observed,
although the stabilization of the peptide is much weaker in this case. 
Again, a markedly reduced $\ku$ is observed, whereas 
the change in $\kf$ is smaller. Therefore, in both the BPTI and GB1 simulations, 
the peptide seems to interact more efficiently with the crowders when 
folded than when unfolded. At the same time,  the peptide-crowder 
interaction is different in character in the BPTI and GB1 cases (see above). 
Note therefore that the folding of the peptide to its native state
entails the formation of both $\beta$-sheet structure and a 
hydrophobic side-chain cluster, which may enhance the interaction 
with GB1 and BPTI, respectively.

\begin{table}[] 
\caption{Folding and unfolding rates of the GB1m3 peptide, $\kf$ and $\ku$, 
in our three simulated systems. The rates are computed from  
the apparent rate constant $k=\kf+\ku$ and the native-state probability, $\Pn$.
The peptide is taken as native if at least three native H bonds are present, and 
$k$ is obtained by fits to the data in Fig.~\ref{fig:autocorr_folding}. Rates are
in units of ($10^6$ MC sweeps)$^{-1}$.}
\begin{center}
\begin{tabular}{lcccc}
\hline\hline
System & $\Pn$ & $k$ & $\kf$ & $\ku$\\
\hline
No crowders	& $0.30\pm0.01$ & $3.8\pm0.3$  & $1.1\pm0.1$ & $2.7\pm0.2$ \\
BPTI	 crowders	& $0.72\pm0.02$ & $1.5\pm0.2$ & $1.1\pm0.1$ & $0.4\pm0.1$ \\
GB1 crowders	& $0.33\pm0.01$ & $2.8\pm0.1$ & $0.9\pm0.1$ & $1.9\pm0.1$ \\
\hline\hline
\end{tabular}
\end{center}
\label{tab:rates}
\end{table}

\section{Discussion and summary}

In this article, we have analyzed the interplay between peptide folding
and peptide-crowder interactions in MC simulations of the GB1m3
peptide with protein crowders, using TICA and MSM techniques.  
A common major advantage of these methods is that they 
can be used to search for key coordinates of complex systems in
an unsupervised manner. We used the simpler TICA 
method to explore the free-energy landscape of the peptide. Using 
a few slow TICA coordinates, it was possible to identify the major 
free-energy minima of the peptide in the presence of the crowders.   

In order to quantitatively analyze the dynamics of the peptide in the 
simulations, we built MSMs. MSMs offer a convenient method
for estimating relaxation times, from the eigenvalues via 
Eq.~\ref{eq:timescales}. However, this method is 
subject to well-known systematic uncertainties. In particular, 
it assumes effectively Markovian dynamics, which, at a given level 
of coarse graining, need not hold 
for small lag times $\ttm$. Unfortunately, in our systems, $\ttm$ 
had to be comparable to the relaxation time in question to keep
the systematic error low. Instead, we therefore estimated
relaxation times by a procedure based on fits to autocorrelation
data for the MSM eigenfunctions. The estimates obtained this way
show essentially no $\ttm$-dependence. This robustness suggests
that the calculated MSM eigenfunctions maintain significant
overlaps with the respective true eigenfunctions down to 
the smallest $\ttm$ values used.   
  
It is, of course, also possible to estimate relaxation times 
from autocorrelation data for other functions than the MSM 
eigenfunctions. However, the autocorrelation of a general 
function is a multi-exponential whose parameters may be  
statistically challenging to determine. The autocorrelation
of an MSM eigenfunction should, by contrast, be close to 
single-exponential over a range of $\tau$, if this eigenfunction
approximates the true eigenfunction sufficiently well (at low 
and high $\tau$, deviations will occur, since the approximation 
is not perfect). The autocorrelations of our MSM eigenfunctions 
showed this behavior, and relaxation times could therefore 
be estimated by single-exponential fits in an  
intermediate range of $\tau$ (where $0.2<C(\tau)<0.8$).
If general functions rather than the MSM eigenfunctions had
been used, our possibilities to estimate relaxation times
would have been much more limited.  

Our simulations further suggest that the GB1m3 peptide 
interacts more efficiently with both BPTI and GB1 
when folded than when unfolded. The addition of either of the
crowders led to a reduced unfolding rate $\ku$, while 
the change in the folding rate $\kf$ was smaller, especially with 
BPTI crowders.

\section*{Supplementary Material}

See supplementary material for illustrations of 
(i) the free energy of GB1m3 with GB1 crowders as
a function of the TIC0 and TIC1 coordinates (Fig.~S1), 
(ii) the preferred GB1m3-GB1 binding modes (Fig.~S2), 
and (iii) the character of the leading MSM  
eigenfunctions in the different systems (Figs. S3--S6).  

\section*{Acknowledgments}
This work was in part supported by the Swedish 
Research Council (Grant no. 621-2014-4522)
and the Swedish strategic research program eSSENCE.
The simulations were performed on resources 
provided by the Swedish National Infrastructure for 
Computing (SNIC) at LUNARC, Lund University, Sweden,
and J\"ulich Supercomputing Centre, Forschungszentrum J\"ulich,
Germany.


%

\end{document}